\pdfoutput=1

\documentclass[aps,prd,twocolumn,showpacs,preprintnumbers,amsmath,amssymb,nofootinbib,superscriptaddress,showkeys]{revtex4-1}

\usepackage{mathtools}
\usepackage{slashed}

\usepackage{float}
\usepackage{epsfig}
\usepackage{graphicx}
\usepackage{latexsym}
\usepackage{longtable}
\usepackage{dcolumn}
\usepackage{braket}
\usepackage[utf8]{inputenc}
\usepackage{soul}

\newcommand\underrel[2]{\mathrel{\mathop{#2}\limits_{#1}}}

\begin{document} 
\hbadness=10000

\title{Baryonic form factors of the pion and kaon in a chiral quark model}

\author{Wojciech Broniowski}
\email{Wojciech.Broniowski@ifj.edu.pl}
\affiliation{H. Niewodnicza\'nski Institute of Nuclear Physics PAN, 31-342 Cracow, Poland}
\affiliation{Institute of Physics, Jan Kochanowski University, 25-406 Kielce, Poland}

\author{Enrique Ruiz Arriola}
\email{earriola@ugr.es}
\affiliation{Departamento de F\'{\i}sica At\'{o}mica, Molecular y Nuclear and Instituto Carlos I de  F{\'\i}sica Te\'orica y Computacional,  Universidad de Granada, E-18071 Granada, Spain}

\author{Pablo Sanchez-Puertas}
\email{psanchez@ifae.es}
\affiliation{Institut de F\'{i}sica d'Altes Energies (IFAE) \& The Barcelona Institute of Science and Technology (BIST), Campus UAB, E-08193 Bellaterra (Barcelona), Spain}

\date{20 July 2022}  

\begin{abstract}
The 
baryonic form factor of charged pions is studied in detail in the Nambu--Jona-Lasinio model with  constituent quarks, where the spontaneously broken chiral symmetry is the key dynamical ingredient guaranteeing the would-be Goldstone  boson nature of the pseudoscalar mesons octet. In general, this form  factor arises when the isospin symmetry is broken, which is the case  if the $u$ and $d$ quark masses are split, as in the real world, or  if the electromagnetic effects were taken into account.  
We obtain estimates for this basic property of the pion resulting from the quark mass splitting for a range of model parameters, and importantly, for different pion masses, going up to the values used in lattice  QCD. We find very stable model results, with the mean square radius of $\pi^+$ in the range $(0.05-0.07~{\rm fm})^2$.  From charge conjugation, the baryonic form factor of $\pi^+$ and $\pi^-$ are  equal and opposite. We also obtain the transverse-coordinate,  relativistically invariant baryonic density of the charged pion. In $\pi^+$, the inner region carries a negative, and the outside -- a positive baryon number density, both cancelling to zero, as obviously the pion carries no net baryon charge. 
We also carry out  an analogous analysis for the kaon, where the effect is much larger  due to the sizable $s$ and $u/d$ quark mass splitting.  We discuss the  prospects of lattice QCD measurements of the baryonic form factors  of charged pions and kaons.  
\end{abstract}

\maketitle

\section{Introduction}
\label{sec:intro}

As it is well known, by construction mesons carry no net baryon charge. However, in a
recent paper~\cite{Sanchez-Puertas:2021eqj} we brought up the largely overlooked 
fact that the {\em baryonic} form factor of charged pions does
not vanish when the isospin symmetry is broken, as is the case when
the $u$ and $d$ quark masses are not equal, or when the
electromagnetic (EM) effects are taken into account.  We carried out several
estimates of the effect based on very different approaches, ranging
from simple quark models to an extraction from the available
$e^+e^-\to \pi^+\pi^-$ data (BaBar~\cite{Aubert:2009ad} and
KLOE~\cite{Aloisio:2004bu,Ambrosino:2008aa,Ambrosino:2010bv,Anastasi:2017eio}),
made with the help of the vector meson dominance (VMD)
involving the $\rho-\omega$ mixing. Our data analysis~\cite{Sanchez-Puertas:2021eqj}  yielded the
following estimate for the baryonic mean squared radius (msr) of
$\pi^+$:
\begin{eqnarray}
 \langle r^2 \rangle_B^{\pi^+} = (0.041(1)~{\rm fm})^2  = 0.0017(1) ~{\rm fm}^2 \, .  \label{eq:data}
\end{eqnarray}
As expected from the weak effect of isospin breaking, this is small compared
to the central value of the accurately known EM radius $\langle r^2 \rangle_Q^{\pi^+} =
(0.659(4)~{\rm fm})^2=0.434(5)~{\rm fm}^2 $~\cite{Zyla:2020zbs}, and at the level of about one third of the
quoted error.

The fact that the charged pion has yet another, hitherto unexplored
form factor corresponding to a conserved current is fundamental and definitely worth further dedicated
studies.  It has a very intuitive physical interpretation in the
coordinate space (in the Breit frame), or in the transverse-coordinate
space~\cite{Burkardt:2000za,Miller:2010nz,Lorce:2020onh}, where
~(\ref{eq:data}) implies that in $\pi^+$ the outer region has a net
baryon density from the excess of the lighter $u$ quark, whereas the
inner region has a net antibaryon density from the heavier $\bar d$
antiquark.  Of course, over the whole space the baryon and antibaryon
densities compensate each other such that the total baryon number of
the pion is zero. For $\pi^-$, the described geometric picture is
opposite, with more antibaryon in the outer region.

The presence of the baryonic form factor of charged pions allows one
for a natural interpretation of the VMD modeling with $\rho-\omega$
mixing, where $\omega$ couples to the baryon
current~\cite{Sanchez-Puertas:2021eqj}. The extraction from the data
in the time-like region, where the mixing becomes most visible,
requires conventional model assumptions regarding the shape of the
higher resonance profiles. An extrapolation to the space-like region
proceeds via a dispersion relation~\cite{Sanchez-Puertas:2021eqj}.

In this paper we focus entirely on the estimates of the non-vanishing
baryonic form factor of pseudoscalar mesons (pions and kaons) based on a chiral quark
model, namely the Nambu--Jona-Lasinio (NJL) model with the
Pauli-Villars (PV) regularization (see~\cite{RuizArriola:2002wr} and
references therein). The model ensures that both pions and kaons emerge as
would-be pseudoscalar Goldstone bosons, thanks to the spontaneous
breakdown of the chiral symmetry. We note that this model has been used to
obtain successful phenomenology for a great variety of soft matrix
elements involving pseudo-Goldstone bosons, hence one may hope it produces
credible results for the baryonic form factor as well. Moreover, in
the model we can easily change parameters. 
In particular, one can increase the pion mass up to 
larger values, such as those employed in some lattice 
QCD simulations, which of course could not be made in an extraction from the
experimental data. One can also increase the $u$ and $d$ mass
splitting, which augments the effect up to the point where it can be
easily observed on the lattice.

The baryonic form factor of charged pions and kaons is proportional to
the splitting of the masses of the constituents that build up the meson.
Compared to~\cite{Sanchez-Puertas:2021eqj}, in the present work we
analyze this splitting more thoroughly, which augments the model
estimate for the baryonic msr of the pion.  We derive analytic expressions,
allowing for a better comprehension of the dependence of the effect on the
kinematic and model parameters. The full details of our calculations
are given in the appendices.

We also analyze the case of the neutral kaons, where (for
structureless quarks as in the NJL model at the leading-$N_c$ level) 
the baryonic form factor is,
up to an overall sign, equal to the EM form factor, for which
experimental and lattice data do exist.
For the case of the pion, where $m_d-m_u$ is tiny compared to other
scales in the model, the baryonic form factor can be evaluated to first
order in $m_d-m_u$. For the kaon, however, all orders in $m_s-m_{u/d}$,
which are substantial, should be kept.  Hence we carry out the calculation
exactly at the one-quark-loop level, which within the NJL model
corresponds to the leading-$N_c$ approximation.

\section{Symmetries and baryonic form factor of the pion \label{sec:sym}}

We begin by reviewing some well-known basic facts, for completeness and to
establish our notation. We then proceed to show that the symmetries do
not preclude a non-zero baryonic form factor of charged pions.  We
recall that in QCD the vector currents corresponding to any flavor
$f=u,d,s,c,b,t$, defined as
\begin{eqnarray}
J_f^\mu(x)= \bar q_f (x) \gamma^\mu q_f (x),  \label{eq:defc}
\end{eqnarray}
are conserved:
\begin{eqnarray}
\partial_{\mu} J_f^\mu(x)= 0, \label{eq:cons}
\end{eqnarray}
with $q_{f}(x)$ denoting a quark field with $N_c=3$ colors (the summation over color is understood).
One introduces the baryon current and the third isospin component of the isovector current as
\begin{equation}
J_B^\mu = \frac{1}{N_c} \sum_f J_f^\mu, \;\;
J_3^\mu = \frac{1}{2} \left (J_u^\mu-J_d^\mu \right). \label{eq:defc2}
\end{equation}
For the considered case of the pion, one can ignore the strangeness and
heavier flavor contributions to matrix elements of $J_B^\mu$, 
as they are strongly suppressed by the Okubo-Zweig-Iizuka (OZI)  
rule and subleading in the large-$N_c$ limit. Hence we take $J_B^\mu = (J_u^\mu+J_d^\mu )/N_c$.
The EM current follows from the Gell-Mann--Nishijima formula,
\begin{eqnarray}
J_{Q}^{\mu} = J_{3}^{\mu} + \frac{1}{2}J_{B}^{\mu}. \label{eq:GN}
\end{eqnarray}
The baryon, the (third component of) isospin,
and the electromagnetic  form factors are defined via matrix elements of the corresponding currents in on-shell pion states of isospin $a=0,+,-$, namely
\begin{equation}
\langle \pi^a({p+q}) \mid J_{B,3,Q}^\mu(0) \mid \pi^a({p})\rangle = (2p^\mu +q^\mu) F^a_{B,3,Q}(t), \label{eq:ffdef}
\end{equation}
with $F_{Q}^a(t)=F_3^a(t)+\frac{1}{2}F_B^a(t)$, in line with Eq.~(\ref{eq:GN}). Here and in the following, $p$ is the momentum 
of the initial pion, $p+q$ of the final pion, and $q^2 \equiv t$. 
We note the following scaling with number of QCD colors: $J_3 \sim 1$ and $J_B \sim 1/N_c$.

Now we pass to symmetries. The currents $J_{B,3}^\mu$ are odd under the charge conjugation $C$, 
whereas the neutral pion is an eigenstate, 
$C |\pi^0\rangle= |\pi^0\rangle$,
which immediately implies 
\begin{eqnarray}
F^{\pi^0}_{B,3}(t)=0 \label{eq:Cpi0}
\end{eqnarray}
identically (at any $t$).
On the other hand, since $C |\pi^\pm\rangle= |\pi^\mp\rangle$, for the charged pions the $C$ conjugation only yields the condition
\begin{eqnarray}
F^{\pi^+}_{B,3,Q}(t) = -F^{\pi^-}_{B,3,Q}(t), \label{eq:Csym}
\end{eqnarray} 
and, of course, no vanishing follows. 

In the case of an exact isospin symmetry, which holds when the light
current quark masses are equal, $m_u=m_d$, and when the small EM
effects are ignored, $G$-parity is also a good symmetry. It is defined
as the charge conjugation followed with the rotation by $\pi$ about the
2-axis in isospin $G=e^{i \pi I_2}C$. All the pions are eigenstates of
$G$, namely $G |\pi^a\rangle = - |\pi^a\rangle$. Since $G$ involves a
flip between $u$ and $d$, $J_3$ becomes even under $G$ (yielding no
constraints), whereas the baryon current remains odd, implying for any
$a$
\begin{eqnarray}
F^{\pi^a}_{B}(t) = 0 \;\;\;\; ({\rm for~an~exact~isospin~symmetry}). \label{eq:Gsym}
\end{eqnarray} 
 
The key point now is that the isospin is only an approximate symmetry
of Nature, broken with $m_d  \neq m_u$ and with EM
interactions. Consequently, $G$-parity is no longer a good symmetry
and the condition~(\ref{eq:Gsym}) need not, and as we explicitly show in~\cite{Sanchez-Puertas:2021eqj} and
in the present paper, does not hold for charged pions in a field theoretical
model with conserved baryon current.

Additional constraints for the form factors at vanishing $t$ follow
from the additivity of charges.
This feature holds in any local field
theory for charges associated with conserved local currents. 
In particular, $F_3^{\pi^+}(t)=1$, which is a sum of isospin charges of
$u$ and $\bar{d}$, analogously $F_3^{\pi^-}(t)=-1$, and
\begin{eqnarray}
F_B^{\pi^\pm}(0)=0, \label{eq:additive}
\end{eqnarray}
which sums to zero the opposite baryon charges of a quark and antiquark. 

To summarize, Eqs.~(\ref{eq:Cpi0}, \ref{eq:Csym}) hold as long as $C$
is a good symmetry (strong and EM interactions),
Eq.~(\ref{eq:additive}) is alway true on general field-theoretic
grounds, whereas Eq.~(\ref{eq:Gsym}) does not in general hold in the
real world, where $m_d > m_u$ and EM interactions are involved.

Although it may seem unusual at first glance, the fact that neutral
particles have a corresponding non-zero charge form factor is not
uncommon.  The neutron carries no electric charge, but has a non-zero
electric form factor, with msr equal to $\langle r^2
\rangle^n_Q=-0.1161(22)~{\rm fm}^2$. As already mentioned at the end of Sec.~\ref{sec:intro}, 
the same is true for neutral kaons.  Even more
unexpectedly, the nucleon possesses a non-vanishing strangeness form
factor, despite being strangeless~\cite{Cohen:1993wn,Forkel:1994yx}.

In this work, similarly as in \cite{Sanchez-Puertas:2021eqj}, we only
explore the {\em charge symmetry breaking} (nomenclature borrowed from
nuclear physics, where $m_n > m_p$) in quark models, i.e. the effects
of $m_d > m_u$. An analysis of EM effects would be much more
involved and extends beyond the scope of the present work.

\section{Chiral quark models at the one-quark-loop level}

We carry out our calculations within the NJL model, where the
(point-like) four-quark interactions lead to dynamical chiral symmetry
breaking, amending quarks with large constituent masses (see
\cite{RuizArriola:2002wr} for a review, and references therein).  At
the leading-$N_c$ level, various observable quantities are evaluated
using one-quark loop.  We stress that in this treatment the pion is
described in a covariant, fully relativistic manner in terms of the
corresponding Bethe-Salpeter equation.  The model is designed for soft
physics, where the virtualities of quarks in the loop are small and
Euclidean,\footnote{Note a recent work~\cite{Miramontes:2021xgn} based on the Bethe-Salpeter equations in the Dyson-Schwinger formalism, where the pion charge form factor is accessible also at physical momenta.} hence a regularization has to be used to cut off the hard
momenta. To preserve the Lorentz, gauge, and chiral symmetries, care
is needed here. A regularization scheme that works, applied in this
work, is based on the PV regularization with two
subtractions~\cite{Schuren:1991sc,RuizArriola:2002wr} (see Appendix
\ref{app:PaVe} for details).

\begin{figure*}[tb]
\centering 
\includegraphics[width=.27\textwidth]{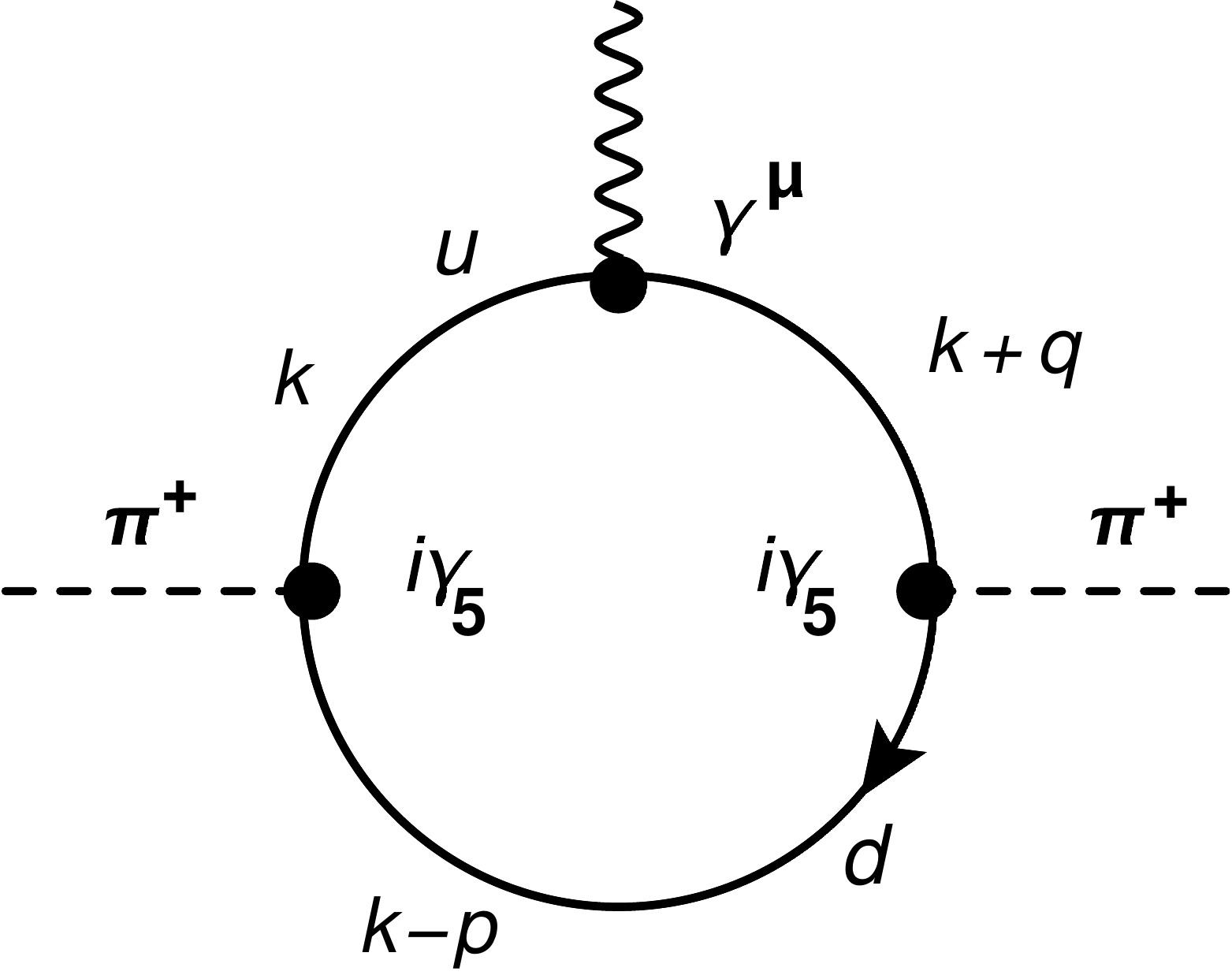} ~~~~~~~ \includegraphics[width=.27\textwidth]{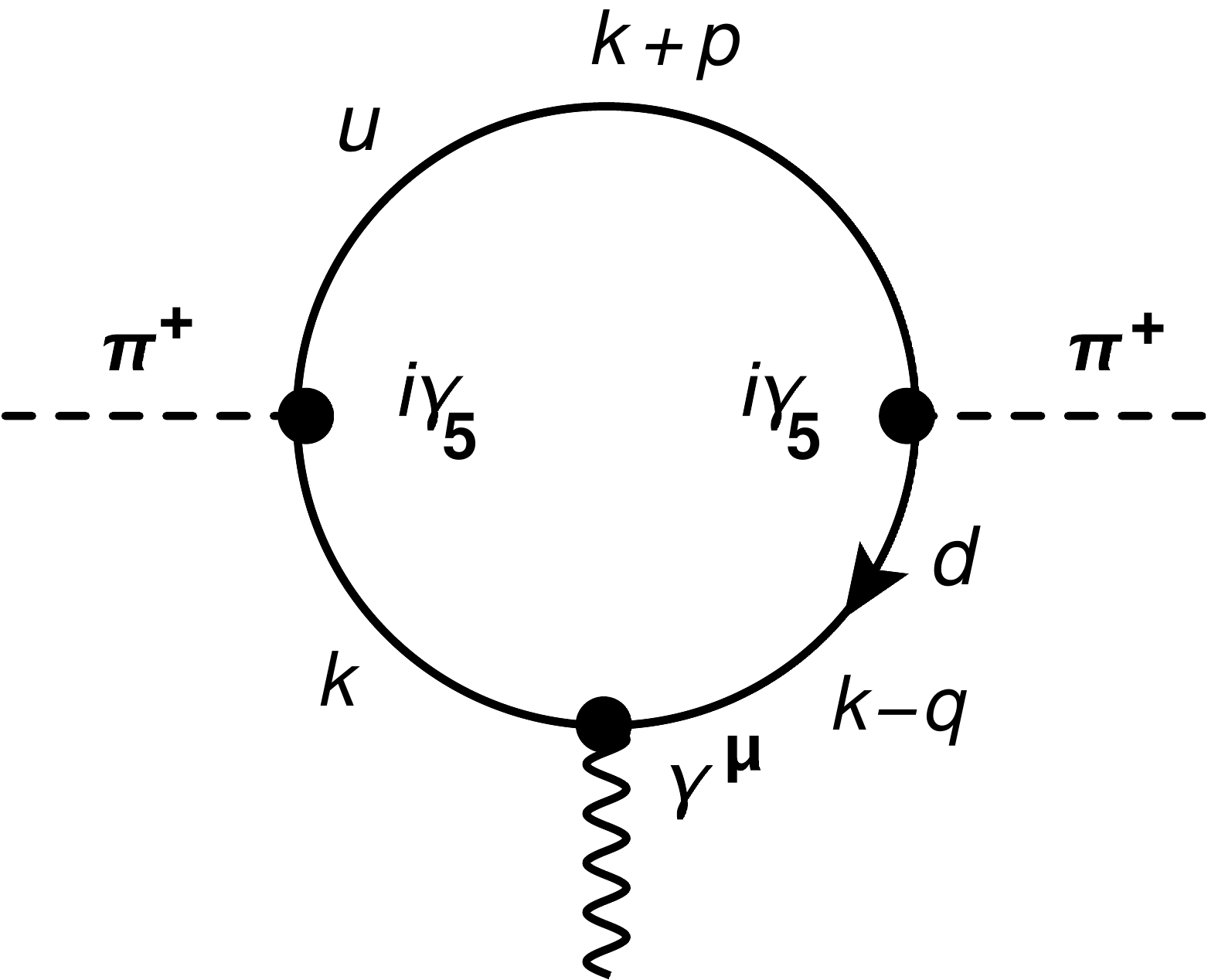}
\caption{\label{fig:feyn} Feynman diagrams for the one-loop evaluation of form factors $F_3$ and $F_B$ for a charged pion  
in chiral quark models. }
\end{figure*}

The approach has been used successfully for a great variety of soft
matrix elements, shedding light also onto such quantities as the parton
distribution functions~\cite{Davidson:1994uv}, parton distribution
amplitudes~\cite{RuizArriola:2002bp}, generalized parton distribution
functions~\cite{Broniowski:2007si}, or the double parton distributions
of the pion~\cite{Courtoy:2019cxq,Broniowski:2019rmu} (see e.g.
\cite{Broniowski:2020jid} for a review).  As the NJL model with
unequal quark masses is not so frequently used (see, however,
\cite{Hutauruk:2016sug,Hutauruk:2018zfk} for the model in the
proper-time regularization), in Appendix~\ref{app:gloss} we provide a
glossary of the NJL model formulas used in our calculations. We also
discuss there the standard strategy of fixing the model parameters
such as the average current quark mass and the PV cut-off.

In the following, we use the short-hand notation for current quark mass average and for the splitting, 
\begin{eqnarray}
m = \frac{m_u+m_d}{2}, \;\;\; \delta = m_d-m_u, \label{eq:delta}
\end{eqnarray}
and for the constituent quarks, correspondingly, 
\begin{eqnarray}
M = \frac{M_u+M_d}{2}, \;\;\; \Delta = M_d-M_u. \label{eq:Delta}
\end{eqnarray}

The one-loop Feynman diagrams used to evaluate the form factors for currents $J_u$ and $J_d$ are presented in Fig.~\ref{fig:feyn}, where 
we have chosen $\pi^+$ for definiteness. The coupling of the pion to the quarks carries the coupling constant $g_{\pi^+u\bar{d}}$ 
discussed in Appendix~\ref{app:gloss}. The expression for the matrix elements corresponding to the two diagrams are 
\begin{widetext}
\begin{eqnarray}
&& \hspace{-4mm} \langle \pi^+(p) \mid J_u^\mu(0) \mid \pi^+(p+q)\rangle = -(\sqrt{2} g_{\pi^+ u \bar{d}})^2 \int \frac{d^4k}{(2\pi)^4} 
{\rm Tr} \left [ \gamma^\mu \frac{i}{\slashed{k}-M_u} \gamma_5 \frac{i}{\slashed{k}-\slashed{p}-M_d} 
      \gamma_5 \frac{i}{\slashed{k}+\slashed{q}-M_u} \right], \nonumber \\
&& \hspace{-4mm}  \langle \pi^+(p) \mid J_d^\mu(0) \mid \pi^+(p+q)\rangle =  - (\sqrt{2} g_{\pi^+ u \bar{d}})^2 \int \frac{d^4k}{(2\pi)^4} 
{\rm Tr} \left [ \gamma^\mu \frac{i}{\slashed{k}-\slashed{q}-M_d} \gamma_5 \frac{i}{\slashed{k}-\slashed{p}{- M_u}} 
      \gamma_5 \frac{i}{\slashed{k}-M_d} \right],  \label{eq:feyn}
\end{eqnarray}
\end{widetext}
with the $+i \epsilon$ prescription in the propagators understood. The trace involves color and Dirac spaces. 
In the adopted convention, the $\sqrt{2}$ factors in the coupling follow from the isospin Clebsch-Gordan coefficients.
The $F_{3,B}$ form factors are then evaluated according to 
Eq.~(\ref{eq:ffdef}) in a standard way. Convenient algebraic Mathematica packages for this task 
are FeynCalc \cite{Shtabovenko:2016sxi,Shtabovenko:2020gxv} 
and Package-X \cite{Patel:2015tea}, which is also capable of evaluating the Passarino-Veltman (PaVe) one loop functions~\cite{Passarino:1978jh} (see
Appendix ~\ref{app:PaVe}) up to three propagators. The result is 
\begin{widetext}
\begin{eqnarray}
F^{\pi^+}_3(t) & =& \frac{2 N_c\pi ^2  g^2_{\pi^+u\bar{d}}}{t-4 m_{\pi }^2} \left [ \left(2 \Delta^2-2 m_{\pi }^2+t\right) B_0^\Lambda\left(t,M_d^2,M_d^2\right)
 -2 \left(\Delta^2+m_{\pi }^2\right) B_0^\Lambda\left(m_{\pi }^2,M_d^2,M_u^2\right) \right . \nonumber  \\
  && + \left . 2 \left(-\Delta^3
   M_d-\Delta M_u \left(\Delta^2-2 m_{\pi }^2+t\right)+m_{\pi }^4\right)
   C_0^\Lambda \left(t,m_{\pi }^2,m_{\pi }^2,M_d^2,M_d^2,M_u^2\right) \right ]  + (u \leftrightarrow d, \; \Delta \to - \Delta), \label{eq:F3}
\end{eqnarray}   
\begin{eqnarray}
\hspace{-10mm} F^{\pi^+}_B(t) & =& \frac{4 \pi ^2  g^2_{\pi^+u\bar{d}}}{t-4 m_{\pi }^2} \left [ 
- \left(2 \Delta^2-2 m_{\pi }^2+t\right) B_0^\Lambda\left(t,M_d^2,M_d^2\right) \right . \nonumber \\
 && \left . -2 \left(-\Delta ^3 M_d-\Delta  M_u \left(\Delta ^2-2 m_{\pi }^2+t\right)+m_{\pi }^4\right) C_0^\Lambda\left(t,m_{\pi }^2,m_{\pi}^2,M_d^2,M_d^2,M_u^2\right) \right ]  - (u \leftrightarrow d, \; \Delta \to - \Delta), \label{eq:FB}
\end{eqnarray}   
\end{widetext}
where $B_0^\Lambda$ and $C_0^\Lambda$ are the PaVe functions in the PV
regularization, see Appendix~\ref{app:PaVe}.  Their explicit forms for
general kinematics and quark masses are analytic, but lengthy. These
exact formulas are used to obtain the results presented in the
following sections.

On general symmetry grounds, as is also apparent from the explicit
expressions in Appendix~\ref{app:gloss}, the quantities $m_\pi^2$ and
$g_{\pi^+u\bar{d}}^2$ appearing in (\ref{eq:F3},\ref{eq:FB}) are even
under the exchange $u \leftrightarrow d$, hence are even functions of
$\Delta$.  Therefore $F_{3}^{\pi^+}$ ($F_{B}^{\pi^+}$) is an even (odd) function of
$\Delta$ and, correspondingly, a series expansion of $F_{3}^{\pi^+}$ ($F_{B}^{\pi^+}$)
involves only even (odd) powers of $\Delta$.

Much simplified formulas for the form factors follow in the chiral
limit of $m_\pi^2=0$ and in the leading order in $\Delta$, which we present for a
better understanding of the following estimates.  To the leading order
in $\Delta$ and in the chiral limit the Goldberger-Treiman relation
holds,
\begin{eqnarray}
g_{\pi^+ u\bar{d}}=\frac{M}{f}, \label{eq:gt}
\end{eqnarray}
where $f=86$~MeV is the value of the pion weak-decay constant,  $F_\pi$, in the chiral limit. 
With these simplifications we arrive at
\begin{eqnarray}
F^{\pi^+}_3(t) &=&\left . 1+\frac{M^2 N_c}{4 \pi ^2 f^2} \left(2 -\sigma\log \left(\frac{\sigma+1}{\sigma-1}\right)\right) \right |_{\rm reg},  
\end{eqnarray}
\begin{eqnarray}
F^{\pi^+}_B(t) &=& \left . \frac{\Delta M^3}{2\pi ^2 f^2 t} \left[ \log^2 \left(\frac{\sigma+1}{\sigma-1}\right)
-\frac{2}{\sigma}  \log \left(\frac{\sigma+1}{\sigma-1}\right)\right] \right |_{\rm reg}, \nonumber \\ 
&& \label{eq:genchir}
\end{eqnarray}
where we have used the short-hand notation
\mbox{$\sigma=\sqrt{1-{4M^2}/{t}}$}, and `reg' denotes the presence
of regularization.  Since $f \sim 1/\sqrt{N_c}$, we verify that
$F_3 \sim 1$ and $F_B \sim 1/N_c$, as stated earlier.

It is instructive to look at the  low-$t$ expansion of the  general formulas (\ref{eq:F3}, \ref{eq:FB}) in the PV 
regularization, up to order $\Delta$ and $m_\pi^2$. In doing so, we also expand to this order the coupling constant:
\begin{eqnarray}
g^2_{\pi^+u\bar{d}} = \frac{M^2}{f^2} \left [ 1-\frac{N_c \Lambda ^4 m_\pi^2}{12 \pi ^2 f M \left(\Lambda ^2+M^2\right)^2}  \right ],
\end{eqnarray}
where here and below the higher order terms are dropped. The result is
\begin{widetext}
\begin{eqnarray}
F^{\pi^+}_3(t) &=& 1+ t \frac{N_c}{24 \pi ^2 f^2} \left[\frac{\Lambda ^4}{\left(\Lambda ^2+M^2\right)^2}+
\frac{\Lambda ^4 \left(\Lambda ^2+3 M^2\right) m_\pi^2 }{5 M^2 \left(\Lambda^2+M^2\right)^3}  
-\frac{N_c \Lambda ^4 m_\pi^2}{12 \pi ^2 f^2 \left(\Lambda ^2+M^2\right)^2} \right] \nonumber  \\
   &\underrel{\Lambda \to \infty}{=}& 1+ t \frac{N_c}{24 \pi ^2 f^2} \left [ 1 -\frac{2M^2}{\Lambda^2}+\frac{m_\pi^2}{5M^2 } 
  -\frac{N_c m_\pi^2}{12 \pi ^2 f^2} \right ]. \label{eq:3exp}
\end{eqnarray}
\begin{eqnarray}
F^{\pi^+}_B(t) &=& t \frac{\Delta}{24 \pi ^2 f^2 M} \left [ \frac{\Lambda ^4 \left(\Lambda ^2+3 M^2\right)}{\left(\Lambda
   ^2+M^2\right)^3}+\frac{4 \Lambda ^4  \left(\Lambda ^4+6 M^4+4 \Lambda ^2 M^2\right) m_\pi^2}{15
   M^2 \left(\Lambda ^2+M^2\right)^4} 
   -\frac{N_c \Lambda ^4 m_\pi^2}{12 \pi ^2 f^2 \left(\Lambda ^2+M^2\right)^2} \right ] \nonumber  \\
   &\underrel{\Lambda \to \infty}{=}&  t \frac{\Delta}{24 \pi ^2 f^2 M} 
   \left [ 1-\frac{3M^4}{\Lambda^4}+\frac{4m_\pi^2}{15M^2 }  -\frac{N_c m_\pi^2}{12 \pi ^2 f^2}  \right ], \label{eq:fbexp}
\end{eqnarray}
\end{widetext}
As typical constituent quark masses are $M\sim 300$~MeV, we infer from
these formulas that the chiral corrections to the slopes are positive
and small, at a level of a few percent.\footnote{Note, however, that our estimates are obtained
  from the one-quark-loop evaluation. Inclusion of pion loops would
  introduce effects relatively suppressed by $1/N_c$, but chirally
  dominant, as is the case of Chiral Perturbation Theory.}  Also, the
results (\ref{eq:3exp}, \ref{eq:fbexp}) are not far from the infinite
cut-off limit, since $\Lambda \sim 800$~MeV. The ratio of the ms radii is
\begin{eqnarray}
\frac{\langle r^2 \rangle_B^{\pi^+}}{\langle r^2 \rangle_3^{\pi^+}} &=&\frac{\Delta \left(\Lambda ^2+3 M^2\right)}{N_c M \left(\Lambda ^2+M^2\right)}
\left [ 1+ \frac{m_\pi^2 \left(\Lambda ^2-3 M^2\right)}{15 M^2 \left(\Lambda ^2+3 M^2\right)}\right ] \nonumber  \\
   &\underrel{\Lambda \to \infty}{=}&  \frac{\Delta}{N_c M} \left [ 1+\frac{2M^2}{\Lambda^2}+\frac{m_\pi^2}{15M^2 }\right ].  \label{eq:ratr}
\end{eqnarray}

We note that numerically, as follows from the fits shown in the
proceeding sections, $\Delta/M$ is at the level of a few percent, and
the higher-order terms in the expansion of $F_B$, starting at
$(\Delta/M)^3$, are completely negligible.

\section{Quark mass splitting \label{sec:split}}

The formulas of the previous section show the {\it a priori} expected
proportionality of $F_B$ to the constituent mass splitting $\Delta$.
Clearly, to make numerical estimates we need the ``physical'' value
for $\Delta$, which is not a trivial matter.  While for the current
quark masses of all flavors, which are QCD parameters, we have
available information from physical processes and perturbation theory,
for the constituent quarks we need to rely on models. This is because
the very notion of a constituent quark has a meaning only within a
specified model, such as the NJL model in our case.

The first issue is the dependence of a constituent quark mass $M_f$ on
the current quark mass $m_f$.  In NJL, the relation following from
the self-consistent gap equation (see Appendix~\ref{app:gloss} and
Fig.~\ref{fig:expa}) is nearly linear  at low $m_f$, namely $M_f\simeq M_f(m_f=0)+\alpha m_f$,
with $\alpha \sim 2$.  So there is no naive
additivity of the constituent and current quark masses\footnote{This is what was inaccurately assumed
  in~\cite{Sanchez-Puertas:2021eqj}, which lowered the NJL estimates
  presented there.} (that would mean $\alpha=1$), which is a 
feedback effect from the quark loop in the gap equation
(see Appendix~\ref{app:gloss}).

\begin{table*}[tb]
\caption{Parameters for the analysis of the pion baryonic form factor.  The uncertainties in $\delta$, $\Delta$, and $\sqrt{\langle r^2 \rangle_B^{\pi^+}}$ follow from Eq.~(\ref{eq:ratpdg}).\label{tab:par}}
\centering
{\small
\begin{tabular}{|rrrr|rrrr|}\hline
$M$ [MeV] & $\Lambda$ [MeV] & $m$ [MeV] & $\delta$ [MeV] &$\Delta$ [MeV]& $m_\pi$ [MeV] & $F_\pi$ [MeV] & $\sqrt{\langle r^2 \rangle_B^{\pi^+}}$ [fm] \\ \hline
300           &  732                       &   0      & 5.0(9)  & 9(2)              &  0                     & 86                    & 0.062(5) \\
300           &  830                       &  7.0    & 5.0(9)  & 11(2)            &  135                 & 93                    & 0.062(5) \\
300           &  960                       &  41.0    & 5.0(9)  & 13(2)          &  400                & 111                  & 0.066(6) \\
280           &  870                       &  6.5    & 4.7(7)  & 11(2)            &  135                 &  93                   & 0.067(5) \\
350          &   770                       &  7.9    & 5.7(1.0)  & 9(1)           &   135               &  93                     & 0.052(4) \\ \hline
\end{tabular}
}
\end{table*}

The model value of $m$ of Eq.~(\ref{eq:delta}) is obtained in the model 
by fitting the physical mass of the pion, $m_\pi$, which in the absence of 
EM effects is $\sim135$~MeV.\footnote{{The splitting $m_{\pi^+}^2 -m_{\pi^0}^2$ is mainly induced by $\mathcal{O}(\alpha_{\rm QED})$ EM
effects~\cite{Donoghue:1996zn}, while the quark mass splitting effect is negligible, as it appears at $\mathcal{O}(\delta^2)$~\cite{Gasser:1982ap}.}} Our result, depending weakly on the adopted value of $M$, is $m \sim 7-8$~MeV. 
This value is about a factor of two larger than the value quoted by the PDG~\cite{Zyla:2020zbs} at 
the scale $\mu=2$~GeV in the $\overline{\rm MS}$ renormalization scheme: 
\begin{eqnarray}
m(\mu = 2~{\rm GeV})=3.45^{+0.55}_{-0.15}~{\rm MeV}. \label{eq:mpdg}
\end{eqnarray}
The effect is due to the running of $m$ with the scale. In perturbative QCD, at leading order (LO) in $\alpha_S$, one has
\begin{eqnarray}
\frac{m_f(\mu)}{m_f(\mu_0)} = \left ( \frac{\alpha_S(\mu)}{\alpha_S(\mu_0)} \right )^{\frac{4}{\beta_0}}, \label{eq:evo}
\end{eqnarray}
where with three flavors $\beta_0=9$, $\alpha_S=4 \pi/[\beta_0 \log(\mu^2 / \Lambda_{\rm QCD}^2)]$, and $\Lambda_{\rm QCD}=226$~MeV.
Let $\mu_0$ denote the quark-model scale, where $m(\mu_0)=7$~MeV (model fit with $M=300$~MeV) or 8~MeV (model fit with $M=350$~MeV). Then, from Eq.~(\ref{eq:evo}) we can infer the value of $\mu_0$, which becomes 
\begin{eqnarray}
&& \mu_0 = 352^{+68}_{-14}~{\rm MeV} \;\;\;\;\; ({\rm for}~m(\mu_0)=7~{\rm MeV}), \nonumber \\
&& \mu_0 = 314^{+43}_{-10}~{\rm MeV} \;\;\;\;\; ({\rm for}~m(\mu_0)=8~{\rm MeV}), \label{eq:esti}
\end{eqnarray}
where the errors reflect the uncertainty in Eq.~(\ref{eq:mpdg}). We thus see that the quark model scale $\mu_0$ is very low.

Quite remarkably, the estimate of Eq.~(\ref{eq:esti}), especially for the case of 8~MeV, is very close to the value obtained with an altogether 
different method, i.e., by using the evolution of the valence quark fraction in the momentum 
sum rule for the pion parton distribution function. There, one finds 
$\mu_0=313^{+20}_{-10}$~MeV~\cite{Davidson:1994uv,Broniowski:2007si}. One could of course argue here that the use of perturbative evolution down to such low scale is questionable at best, but the fact that the quark model scale is very low touches upon the essence of effective chiral quark models, where there are no explicit gluon degrees of freedom. 

Actually, what we need in the present task is not a precise value of $\mu_0$, but only the fact that the $\overline{MS}$ evolution prescription preserves the ratios of the current quark masses of different flavor, which are independent of the scale:
\begin{eqnarray}
\frac{m_f(\mu)}{m_{f'}(\mu)} = \frac{m_f(\mu_0)}{m_{f'}(\mu_0)}. \label{eq:rat}
\end{eqnarray}
Therefore the ratio $m_d/m_u(\mu_0)$ which we should use in the model is exactly the same as the PDG~\cite{Zyla:2020zbs} value at $\mu=2$~{\rm GeV}:
\begin{eqnarray}
\frac{m_u}{m_d}=0.47^{+0.06}_{-0.07}. \label{eq:ratpdg}
\end{eqnarray}

We can make another independent estimate of the the splitting $\delta$
based on the difference of the $K^0$ and $K^+$ masses. First, one needs
to subtract the EM effects, which contribute $2.6$~MeV more to $K^+$ than to $K^0$~\cite{Donoghue:1996zn}. 
We thus need to adjust in the model
the values of $m_d$ and $m_u$ such that the resulting splitting between $K^0$
and $K^+$ masses is the physical value plus the EM correction,
$3.9+2.6=6.5$~[MeV]. With $m=7$~MeV and $m_s=181$~MeV (needed to fit
the physical value of $m_K$) we get in the model $m_u/m_d\simeq 0.47$, a value in
the center of~(\ref{eq:ratpdg}), whereas with $m=7$~MeV we find
$m_u/m_d\simeq 0.52$, within the error of~(\ref{eq:ratpdg}). To
summarize, in our numerical studies presented below we will use the PDG
ratio (\ref{eq:ratpdg}) as a credible estimate.

Finally, speaking of the QCD evolution, we recall that form factors
corresponding to conserved currents are scale independent.  Therefore
both the charge and the baryonic form factors of the pion obtained at a
given scale (here, at the the quark model scale), are universal. Yet,
to carry out the calculation, we need to know the pertinent quantities,  such as $m$ or $\delta$,
at the scale where the calculation is made.

\section{Results for the pion \label{sec:pion}}

\begin{figure*}[tbp]
\centering 
\includegraphics[width=.463\textwidth]{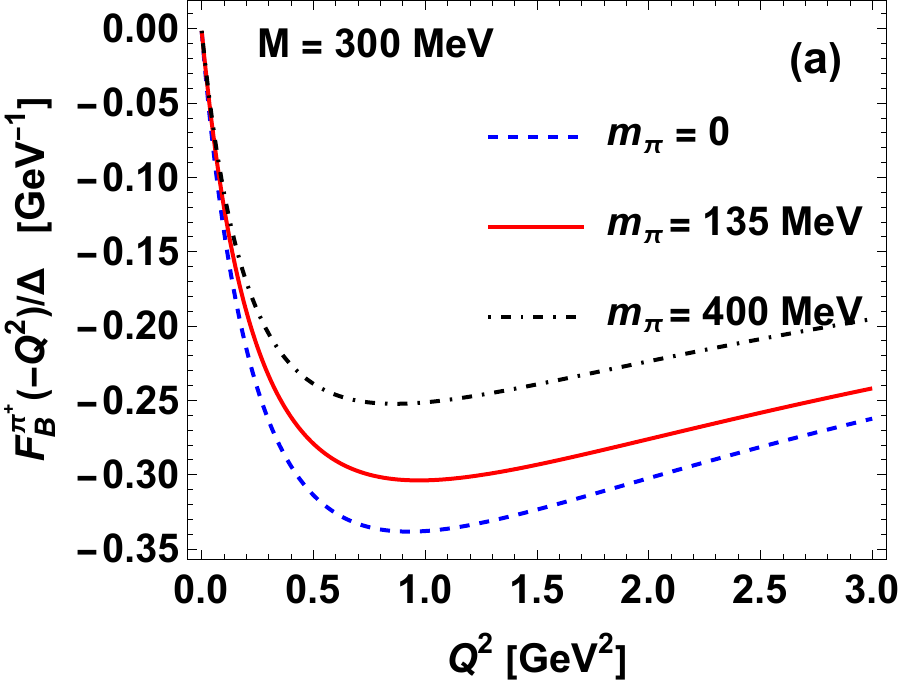}\hfill \includegraphics[width=.47\textwidth]{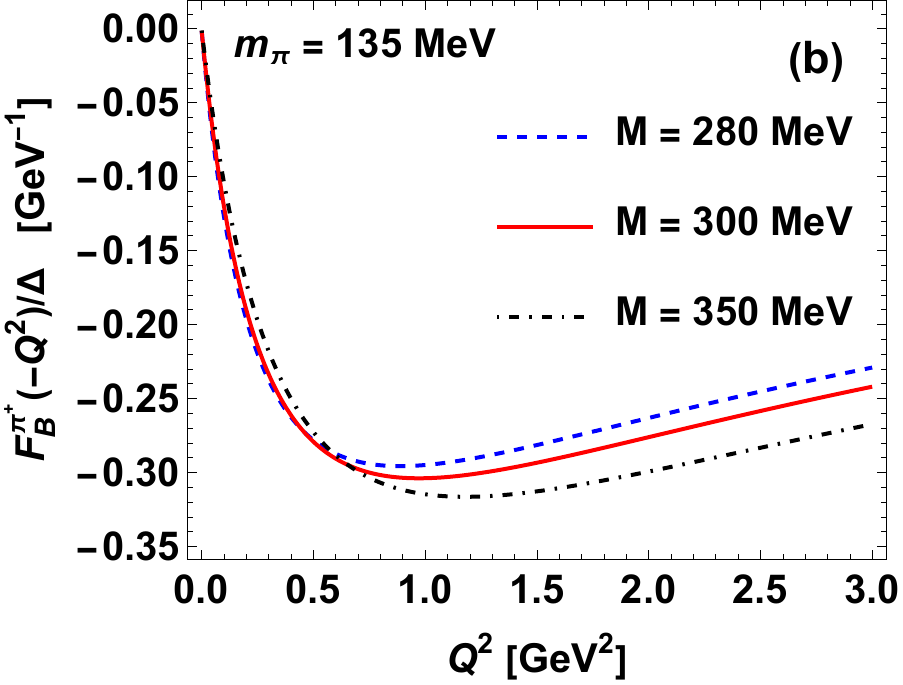} 
\vspace{-1mm}
\caption{\label{fig:mpi} Baryonic form factor of $\pi^+$ divided by $\Delta=M_d-M_u$, obtained from the NJL model with PV regularization.
Parameters of the calculations are listed in Table~\ref{tab:par}. Panel~(a) compares the results for different values of $m_\pi$ at a fixed constituent quark mass $M$, whereas panel~(b) compares the results for different values of $M$ at a fixed $m_\pi$.}
\end{figure*}

The parameters used in our estimates for the pion are collected in Table~\ref{tab:par}. We study the dependence on $m_\pi$ (the first three rows of the table), from the chiral limit, through the physical mass (without EM effects) of 135~MeV, up to a large value of 400~MeV, typical in some lattice QCD simulations.
For the first three rows of Table~\ref{tab:par} the value of $M$ is fixed at 300~MeV, while $\Lambda$ and $m$ are fitted to the values of $m_\pi$ and $F_\pi$ (for the case $m_\pi=400$~MeV we take $F_\pi=110$~MeV, which is in the ball park of the range given in~\cite{RBC:2010qam}).\footnote{{One can also use here the NLO $\chi$PT that leads to similar numbers: 
$F_{\pi}=f [ 1 -2\mu_{\pi} -\mu_K +\frac{4m_{\pi}^2}{f^2}(L_4^r +L_5^r)  +\frac{8m_{K}^2}{f^2}L_4^r ]$, 
with $\mu_P = m_P^2/(32\pi^2 f^2)\ln(m_P^2/\mu^2)$ and $L_{4(5)}^r=-0.3(1.4)\times 10^{-3}$ at $\mu=0.77$~GeV.}} We note that the values of the baryonic msr are very stable, which simply reflects the fact that $\Delta$ changes very little. 
Rows 2, 4, and 5 compare the results at fixed $m_\pi=135$~MeV, but for different values of $M$. Here we note some weak dependence of the values of msr, attributed to different values of $\Delta$ and $M$.

In Fig.~\ref{fig:mpi} we show the dependence of the baryonic form
factor of $\pi^+$ in the space-like domain of $t \le 0$, evaluated
according to Eq.~(\ref{eq:FB}) with the parameters of
Table~\ref{tab:par}. To be less sensitive to the value of $\Delta$, we
plot $F^{\pi^+}_B(t)/\Delta$. The left panel shows the dependence on
$m_\pi$ and the right panel the dependence on $M$. We note very
stable results, in particular at low $-t$. One should bear in mind that
NJL is, by construction, a low-energy model, hence the results at $-t > \Lambda^2$ need not be credible.

Having the form factor in the momentum space, one may construct the transverse density~\cite{Burkardt:2000za,Miller:2010nz,Lorce:2020onh} via a Fourier-Bessel transformation,
\begin{eqnarray}
2\pi b \, \rho_B^{\pi^+}(b) = \int_0^\infty \! dQ \, Qb\, F_B^{\pi^+}(-Q^2) J_0(Q b). \label{eq:bessel}
\end{eqnarray}
This quantity is boost-invariant, hence free of
ambiguities~\cite{Jaffe:2020ebz,Soper:1976jc} present in the popular
Breit-frame density in the radial coordinate. The result for $\pi^+$
is shown in the left panel of Fig.~\ref{fig:rhob}, solid line.  We
note the intuitive mechanistic interpretation, announced in the
introduction. In $\pi^+$, the heavier $\bar{d}$ quark has a more
compact distribution than the lighter $u$ quark, hence at low $b$
there is an excess of antibaryon charge density. At large $b$ the
situation is opposite, such that the constraint of the vanishing total
baryon number, or $\int db \, b \, \rho_B^{\pi^+}(b)=0$, is satisfied.

For $\pi^-$, the corresponding plots of Figs.~\ref{fig:mpi} and
\ref{fig:rhob} are equal and opposite.

Using the relations
\begin{eqnarray}
\langle r^2 \rangle^{\pi^+}_3 &=& \tfrac{1}{2} \langle r^2 \rangle^{\pi^+}_u+\tfrac{1}{2} \langle r^2 \rangle^{\pi^+}_{\bar{d}}, \nonumber \\
\langle r^2 \rangle^{\pi^+}_B &=& \tfrac{1}{3} \langle r^2 \rangle^{\pi^+}_u-\tfrac{1}{3} \langle r^2 \rangle^{\pi^+}_{\bar{d}}, \label{eq:relr}
\end{eqnarray}
we find in our model, with the physical pion mass and $M=300$~MeV, the following msr of the $u$ and $\bar{d}$ constituent quarks in $\pi^+$:
\begin{eqnarray}
\langle r^2 \rangle^{\pi^+}_u &=& 0.273(1)~{\rm fm}^2 = (0.523(1)~{\rm fm})^2, \nonumber \\
\langle r^2 \rangle^{\pi^+}_{\bar{d}} &=& 0.262(1)~{\rm fm}^2 = (0.511(1)~{\rm fm})^2, \label{eq:est}
\end{eqnarray}
with the error reflecting the uncertainty in $\delta$.
This shows the advocated mechanistic feature that the heavier $\bar{d}$ quark has a more compact distribution. In $\pi^-$, the above numbers hold with the replacement $u \to \bar{u}$ and $\bar{d} \to d$

\section{Results for the kaon \label{sec:kaon}}

The case of the kaon is obtained directly from the pion with simple
flavor substitutions. The case considered up to now has been $\pi^+ =
u\bar{d}$. We can pass to $K^+=u\bar{s}$ replacing $d\to s$, and to
$K^0=d\bar{s}$ replacing $u \to d$ and $\bar{d}\to \bar{s}$. All the previously
derived formulas then hold.\footnote{The symmetry arguments of Sec.~\ref{sec:sym} hold for the SU(3) case with the replacement of the I-spin with U-spin or V-spin.}  The parameters used here are
those of the second row of Table~\ref{tab:par}, supplemented with
the current strange quark mass at the scale $\mu_0$ equal to
$m_s=180$~MeV, which fits the kaon mass. Note that the
scale-independent ratio obtained that way, $m_s/m=25.7$, is merely two standard deviations away
from the PDG~\cite{Zyla:2020zbs} value of $27.3^{+0.7}_{-1.3}$.
The model in the strange sector is, however, not perfect, as the value of
the kaon decay constant is $F_K=100$~MeV, with experiment giving
$110(1)$~MeV.
The issue may be improved by introducing more interaction
terms in the NJL Lagrangian (see,
e.g.,~\cite{Bernard:1987gw,Korpa:1989pp,Osipov:2006ns}), but for our
present exploratory study this problem is not critical.

\begin{figure*}[tb]
\centering 
\includegraphics[width=.449\textwidth]{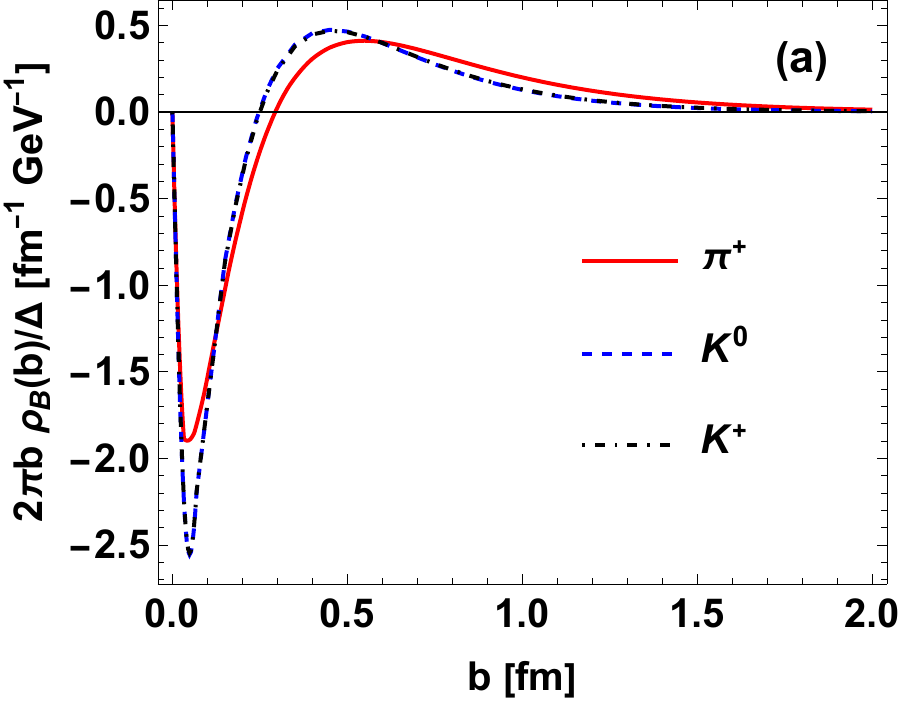} \hfill \includegraphics[width=.47\textwidth]{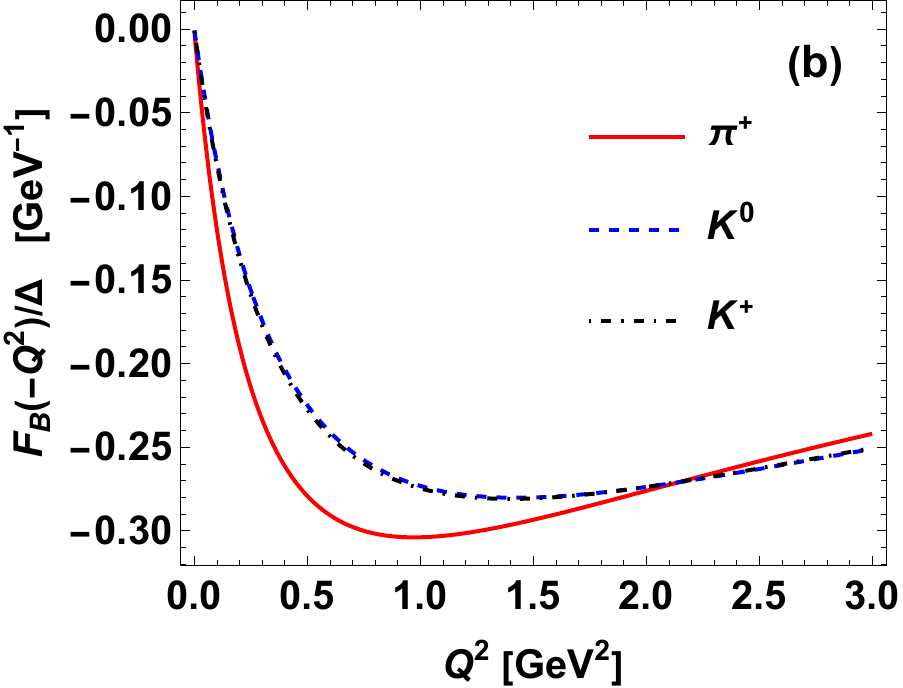}
\vspace{-1mm}
\caption{\label{fig:rhob} The transverse baryonic density (a) and the baryonic form factor (b) for $\pi^+$, $K^0$, and $K^+$, divided by, correspondingly, $\Delta=M_d-M_u$, $\Delta=M_s-M_d$, and $\Delta=M_s-M_u$. The curves for $K^0$ and $K^+$ nearly overlap.}
\end{figure*}

The case of the baryonic form factor of the kaon is compared to the
pion in the right panel of Fig.~\ref{fig:rhob}. The form factor in
the figure is scaled with $\Delta=M_s-M_u$ for the case of $K^+$ and
$\Delta=M_s-M_d$ for the case of $K^0$, whereas for the pion
$\Delta=M_d-M_u$. We note that the behavior of the kaon is similar to
the pion. The scaled curves for $K^+$ and $K^0$ practically coincide.

The left panel of Fig.~\ref{fig:rhob} shows he corresponding
transverse densities. We note that the kaon curve is more compact,
crossing zero at lower $b$. Similarly to the pion case, this feature
is also in accordance to the ``mechanistic'' interpretation mentioned in the Introduction, as the
constituent mass difference is larger in the kaon than in the pion.

The baryonic msr from the model are
\begin{eqnarray}
 \langle r^2 \rangle_B^{K^+} &=&(0.24(1)~{\rm fm})^2=0.056(1)~{\rm fm}^2, \label{eq:kmsr} \\
 \langle r^2 \rangle_B^{K^0} &=&(0.23(1)~{\rm fm})^2=0.052(1)~{\rm fm}^2. \nonumber 
\end{eqnarray}
For structureless quarks, as in NJL in the large-$N_c$ approximation,
\begin{eqnarray}
\langle r^2 \rangle_B^{K^0}=-\langle r^2 \rangle_Q^{K^0}, \label{eq:Kid}
\end{eqnarray}
because the baryon number and charge of the $s$ and $d$ quarks are equal and opposite. PDG~\cite{Zyla:2020zbs} 
quotes $\langle r^2 \rangle^{K^0 }_Q =-(0.28(2)~{\rm fm})^2=-0.077(10)~{\rm fm}^2$, 2.5 standard deviations larger than the value in~(\ref{eq:kmsr}).
We note that the lattice QCD simulations~\cite{Aoki:2015pba} yield $\langle r^2 \rangle_Q^{K^0}=-0.055(15)~{\rm fm}^2$, in agreement with the estimate~(\ref{eq:kmsr}) via Eq.~(\ref{eq:Kid}).

The msr of the constituent quarks, analogous to Eq.~(\ref{eq:est}), are
\begin{eqnarray}
\langle r^2 \rangle^{K^0}_d &=& 0.283(1)~{\rm fm}^2 = (0.532(1)~{\rm fm})^2, \nonumber \\
\langle r^2 \rangle^{K^0}_{\bar{s}} &=& 0.127(1)~{\rm fm}^2 = (0.356(1)~{\rm fm})^2, \nonumber \\
\langle r^2 \rangle^{K^+}_u &=& 0.295(1)~{\rm fm}^2 = (0.543(1)~{\rm fm})^2, \nonumber \\
\langle r^2 \rangle^{K^+}_{\bar{s}} &=& 0.127(1)~{\rm fm}^2 = (0.356(1)~{\rm fm})^2, \label{eq:estK}
\end{eqnarray}
with same numbers holding for $\overline{K}^0$ and $K^-$ upon the replacement of quarks into antiquarks. Again, we note the mechanistic feature of the $\bar{s}$ being significantly more compact, and the distribution of $d$ in $K^0$ more compact than the distribution of $u$ in $K^+$. 

\section{Conclusions}

As we can see from Table~\ref{tab:par}, the NJL model predictions for the baryonic msr of the pion are, taking into account various model parameters,
\begin{eqnarray}
\langle r^2 \rangle^{\pi^+}_B=(0.06(1)~{\rm fm})^2=0.004(1)~{\rm fm}^2. \label{eq:NJLest} 
\end{eqnarray}
This is about a factor of {2} higher than the value extracted 
from the data,~(\ref{eq:data}). Possible reasons for this discrepancy are as follows: 

First, the NJL calculation takes into account only the one-quark-loop
contribution, which is leading in $N_c$, but does not include the
formally suppressed but potentially large chiral loops. For the case
of the isospin msr of the pion, chiral loops contribute about 20\% of
the total result and are necessary to reproduce the data. In the
present case we may expect a similar order on the effect.

Second, the extraction from the data~(\ref{eq:data}) includes EM effects
dressing the pion vertex (although the experimental procedure gets
rid of the EM interactions in the initial and final states). These
effects are not present in our calculation. 

Finally, we comment on the lattice QCD prospects of measuring the
baryonic msr of the pion and kaon.  Comparing our numbers to the
accuracy of the recent lattice QCD calculations of the charge form
factor with physical quark masses, $\langle r^2 \rangle_Q^{\pi} =
(0.648(15)~{\rm fm})^2 =0.42(2)~{\rm fm}^2$~\cite{Gao:2021xsm} and
$\langle r^2 \rangle_Q^{\pi} =0.430(5)(13)~{\rm fm}^2$~\cite{Wang:2020nbf}, we remark that our 
effect~(\ref{eq:data}) is an order of magnitude smaller, and the estimate~(\ref{eq:NJLest}) 
-- a factor of 5 smaller than the lattice accuracy quoted above.  However, on the
lattice one may try to increase the value of the quark mass splitting up
to the point where the signal is strong enough, and then extrapolate
down to the physical point. It would be very interesting to see if with the present 
accuracy such a determination is possible.

For the case of the baryonic msr of the kaon~(\ref{eq:kmsr}),
discussed in Sec.~\ref{sec:kaon}, the baryonic msr of $K^0$ may be
considered to have already been measured on the lattice~\cite{Aoki:2015pba}. It satisfies
identity~(\ref{eq:Kid}), as no disconnected contributions have been accounted for in the simulations, which are estimated to be negligible.

Finally, we wish to underline that one may use different models (all
one needs is current conservation and the ability to model the pion or kaon)
to assess the size of the baryonic form factor. It would
certainly be very interesting to have such independent estimates 
{to confront them with the result (\ref{eq:data}) extracted from the experiment.

\appendix

\section{Regularized one-loop functions \label{app:PaVe}}

All our results involve one loop functions, for which we use the Passarino-Veltman~\cite{Passarino:1978jh} convention 
(the $+i\varepsilon$ prescription in the denominators is understood):
\begin{widetext}
\begin{eqnarray}
i \pi^2 A_0(M_1) &=&\int d^4k \frac{1}{k^2-M_1^2}, \label{eq:pave} \\
i \pi^2 B_0(p^2,M_1,M_2) &=&\int d^4k \frac{1}{[(k+p)^2-M_1^2][k^2-M_2^2]}, \nonumber \\
i \pi^2 C_0(q^2,p^2,(p+q)^2,M_1,M_2,M_3) &=&\int d^4k \frac{1}{[(k+p+q)^2-M_1^2][(k+p)^2-M_2^2][k^2-M_3^2]}. \nonumber
\end{eqnarray} 
\end{widetext}
The $A_0$ and $B_0$ functions are divergent and require regularization. The $C_0$ function is convergent, however, it still should be regularized to dispose of the high-momentum contributions in the quark-loop, which should not enter the low-energy model. 
The applied PV regularization prescription with 
two subtractions~\cite{Schuren:1991sc,RuizArriola:2002wr} amounts to the replacement 
\begin{eqnarray}
F^\Lambda(\{M_i^2\}) &=& F(\{M_i^2\}) - F(\{ M_i^2+\Lambda^2 \} ) \nonumber \\
&+& \Lambda^2 \frac{dF(\{M_i^2+\Lambda^2\})}{d\Lambda^2},
\end{eqnarray}
where $F$ is a the PaVe function (with arguments other than the masses suppressed). The chosen regularization is consistent with the symmetry requirements~\cite{Schuren:1991sc,RuizArriola:2002wr}.

Explicitly 
\begin{eqnarray}
A_0^\Lambda(M_f)=-\frac{M_f^2 \log \left( \frac{M_f^2}{\Lambda^2+M_f^2}\right)+\Lambda^2}{16 \pi ^4}. \label{eq:A0PV}
\end{eqnarray}
Analytic but lengthy formulas for $B_0^\Lambda$ and $C_0^\Lambda$ can be obtained.

\section{Basics of the Nambu--Jona-Lasinio model \label{app:gloss}}

This Appendix presents for completeness the standard NJL formulas in the adopted notation and for the general 
case of unequal $u$ and $d$ quark masses, needed in our analysis. For explanations and the physics discussion the reader is referred to~\cite{RuizArriola:2002wr}.

The gap equation for each flavor $f$ has the form 
\begin{eqnarray}
M_f=m_f-4\pi^2 G M_f N_c A_0^\Lambda(M_f), \label{eq:gap}
\end{eqnarray}
where $G$ is the NJL four-quark coupling constant (independent of flavor or the value of $m_f$). The quark condensate for a single flavor is 
\begin{eqnarray}
\langle \bar{q}_f q_f \rangle = 4\pi^2 M_f N_c A_0^\Lambda(M_f) = - \frac{M_f-m_f}{G}. \label{eq:qcond}
\end{eqnarray} 
The mass of $\pi^\pm$, denoted as $m_\pi^2$, is a root in $p^2$ of the denominator of the pion propagator, following from the Bethe-Salpeter equation,
\begin{eqnarray}
4 \pi ^2 \left(p^2 - \Delta^2\right) N_c B_0^\Lambda(p^2,M_u,M_d)=\frac{1}{G} \left ( \frac{m_u}{M_u}+\frac{m_d}{M_d} \right ), \nonumber \\ \label{eq:pimass}
 \end{eqnarray}  
The coupling constant of a charged pion to the $u$ and $d$ quarks is obtained from the residue of the Bethe-Salpeter amplitude at the pion pole,
\begin{eqnarray}
&& \frac{1}{g^2_{\pi^+ u \bar{d}}} = \frac{1}{g^2_{\pi^- d \bar{u}}} \label{eq:gpi} \\ 
&& =4 \pi^2 \left . \frac{d}{dp^2}\left [
(p^2 - \Delta^2) N_c  B_0^\Lambda(p^2,M_u,M_d) \right ] \right |_{p^2=m_\pi^2}. \nonumber
\end{eqnarray}
The pion weak decay constant is 
\begin{eqnarray}
F_\pi = \frac{2 \pi ^2  g_{\pi^+ u \bar{d}} N_c}{m_\pi^2} 
&& \left [ 2 (m_\pi^2-\Delta^2) M B^\Lambda_0 (m_\pi^2,M_u,M_d) \right . \nonumber \\
&& \left . -\Delta \left(A_0(M_u)-A_0(M_d) \right ) \right ].
\end{eqnarray}
In the limit of $m_u=m_d=0$ the Goldberger-Treiman relation~(\ref{eq:gt}) holds.

\section{Fixing the model parameters and $\Delta$ \label{app:param}}

The model has four parameters: $G$, $m_u$, $m_d$, and $\Lambda$, which with the expressions from the previous Appendix can be traded for 
$M$, $\Delta$, $m_\pi$, and $F_\pi$. Fitting $m_\pi$ and $F_\pi$ to their physical values leaves two parameters: $M$ and $\Delta$. As usual in the NJL studies, we keep $M$ free, whereas $\Delta$ can be related to the splitting of the {\em current} quark masses, $m_d-m_u$.

\begin{figure}[t]
\centering 
\includegraphics[width=.4\textwidth]{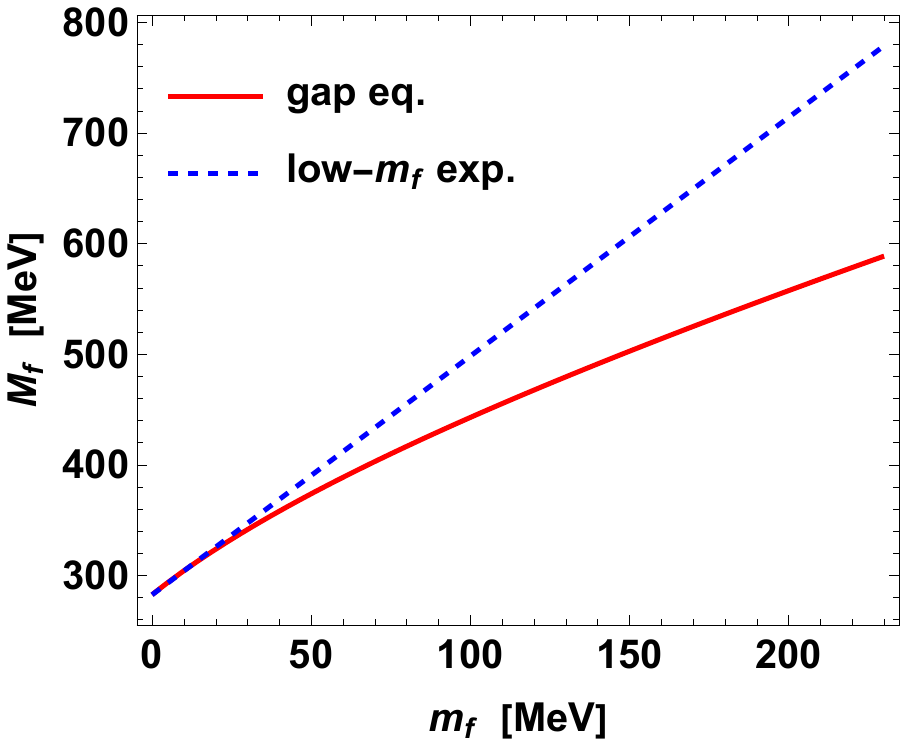} 
\vspace{-2mm}
\caption{\label{fig:expa} Dependence of the constituent quark mass, $M_f$, on the current mass, $m_f$, in the NJL model with the PV regularization.}
\end{figure}

From the gap equation~(\ref{eq:gap}), obviously, $M_f$ is a function of $m_f$. We discuss this issue at a greater length, as it is not 
covered in detail in the literature, while we need it for the determination of the mass splitting $\Delta$ to be used in our estimates.  
Using Eq.~(\ref{eq:qcond}) in Eq.~(\ref{eq:gap}) we arrive at 
\begin{eqnarray}
M_f(m_f)=  M_f(0) \frac{\langle\bar{q}_f q_f \rangle(m_f)}{\langle \bar{q}_f q_f \rangle(0)}  + m_f. \label{eq:Mf}
\end{eqnarray}
At small $m_f$ we may expand the quark condensate, 
\begin{eqnarray}
 \langle\bar{q}_f q_f \rangle(m_f) &=& \langle\bar{q}_f q_f \rangle(0) +\chi_{f} m_f +{\cal O}(m_f^2), \nonumber \\
\chi_{f}&=&\langle\bar{q}_f q_f \rangle',
\end{eqnarray} 
where $\chi_f$ is the quark mass (or scalar) susceptibility in the chiral limit, and the prime denotes a derivative with respect to $m_f$ at $m_f=0$,
Therefore
\begin{eqnarray}
M_f(m_f)&=&M_f(0)+\left [ 1 +  M_f(0) \frac{\chi_f}{\langle\bar{q}_f q_f \rangle} \right ] m_f +{\cal O}(m_f^2) \nonumber \\
 &=&M_f(0)+\alpha m_f +{\cal O}(m_f^2), 
\end{eqnarray}
and in general $\alpha \neq 1$, in contrast to what one might naively assume. The above formula holds for a generic approach with the gap equation.

Explicitly, in the NJL model with PV regularization we find, by expanding Eq.~(\ref{eq:gap}),  
\begin{eqnarray}
&& M'_f(0) m_f = m_f  - \\
&&  \left . 4\pi^2 G N_c \frac{d}{dM_f} M_f A_0^\Lambda(M_f) \right |_{M_f=M_f(0)} M'_f(0) m_f +\dots, \nonumber
\end{eqnarray} 
from where we can evaluate $M'_f(0)$ and obtain
\begin{eqnarray}
&& M_f(m_f)=M_f(0)+  \label{eq:Mexp} \\
&& \frac{m_f}{1+\left . 4\pi^2 G N_c \frac{d}{dM_f} M_f A_0^\Lambda(M_f) \right |_{M_f=M_f(0)}} + {\cal O}(m_f^2). \nonumber
\end{eqnarray}
Eliminating $G$ from Eq.~(\ref{eq:gap}) at $m_f=0$ we get
\begin{eqnarray}
M_f(m_f)=M_f(0) \!-\! \frac{ A^\Lambda_0[M_f(0)] \, m_f }{\left . M_f \frac{d}{dM_f} A_0^\Lambda(M_f) \right |_{M_f=M_f(0)}} \!+ \!{\cal O}(m_f^2). \nonumber \\  \label{eq:Mexp}
\end{eqnarray}
With the typically used parameters, the slope parameter is $\alpha \sim 2$.

Figure~\ref{fig:expa} shows that the small-$m_f$ expansion works well in the range of the light current quark masses, but not for the higher 
values of the strange quark.
Asymptotically, at very large $m_f$ (and, therefore, large $M_f$) we have $A_0^\Lambda(M_f) \sim \Lambda^4/(32\pi^4 M_f)^2$ and
\begin{eqnarray}
M_f=m_f+\frac{G N_c \Lambda^4}{8\pi^2 m_f} + {\cal O}(1/m_f^2).
\end{eqnarray}
Thus, asymptotically, $M_f/m_f \to 1$.

\bigskip

\acknowledgments

Supported by the Polish National Science Centre grant
2018/31/B/ST2/01022 (WB), European H2020 MSCA-COFUND grant 754510
and H2020-INFRAIA-2018-1 grant 824093, by the Spanish MINECO grants
FPA2017-86989-P and SEV-2016-0588, Generalitat de Catalunya grant
2017SGR1069 (PSP), the Spanish MINECO and European FEDER funds grant
and Project No. PID2020–114767 GB-I00 funded by
MCIN/AEI/10.13039/501100011\-033, and by the Junta de Andaluc{\'i}a grant
FQM-225 (ERA).

\bibliography{refs-barpi}

\end{document}